\documentclass[preprintnumbers,amsmath,amssymb,showkeys]{revtex4} 

\usepackage{graphicx}
\usepackage{dcolumn}
\usepackage{bm}

\begin{document}

\preprint{Gen. Relativ. Grav./ DOI: 10.1007/s10714-012-1406-8}

\title{Space-time inhomogeneity, anisotropy and gravitational collapse}

\author{R. Sharma}
\email{rsharma@iucaa.ernet.in}
\affiliation{Department of Physics, P. D. Women's College, Jalpaiguri 735101, India.}

\author{R. Tikekar}
\email{tikekar@gmail.com}
\affiliation{Formerly at: Department of Mathematics, Sardar Patel University, Vallabh Vidyanagar, Gujarat, India.}
          
\date{\today}

\begin{abstract}
We investigate the evolution of non-adiabatic collapse of a shear-free spherically symmetric stellar configuration with anisotropic stresses accompanied with radial heat flux. The collapse begins from a curvature singularity with infinite mass and size on an inhomogeneous space-time background. The collapse is found to proceed without formation of an even horizon to singularity when the collapsing configuration radiates all its mass energy. The impact of inhomogeneity on various parameters of the collapsing stellar configuration is examined in some specific space-time backgrounds.
\end{abstract}

\keywords{Gravitational collapse; Radiating star; Einstein's field equations; Space-time inhomogeneity.}

\maketitle

\section{Introduction}
One of the most fundamental problems in general relativity is the construction of realistic models describing various evolutionary stages of a star collapsing under its own gravity. When a star exhausts all its thermonuclear fuel, it cannot withstand the gravitational pull and starts collapsing. As it contracts appreciably, the density increases and at a sufficiently high density it produces non-thermal pressure via degenerate fermions and particle interactions to support it against further collapse and becomes a `compact star'. Stable stellar configurations like neutron stars are the end products of such radiative collapse processes. A massive star, however, can not come to a stable stage by such processes. In the absence of any mechanism that can withstand the gravitational pull, the general relativistic prediction is that such a collapse must terminate into a space-time singularity. In view of the Cosmic Censorship Conjecture(CCC), such a singularity must be covered within its event horizon of gravity\cite{Joshibook}. However, there are several counter examples where a naked singularity is more likely to be formed. In fact, there is yet no established theory available governing the formation of either a black hole or a naked singularity (for a recent review see \cite{Joshi01} and references therein). The physics of evolving dynamical systems, therefore, continues to generate a great deal of interest in various fields of astrophysics and cosmology even today.

To understand the nature of collapse of a self gravitating body, one needs to provide an accurate description of the exterior and interior space-times of the collapsing body. It is also necessary to find out the appropriate boundary conditions joining the two regions. The theoretical understanding of gravitational collapse was first initiated by Oppenheimer and Snyder\cite{OppenS} who considered the contraction of a highly idealized spherically symmetric dust cloud. The exterior space-time of the collapsing dust cloud was described by the Schwarzschild metric and the interior space-time was represented by a Friedman-like solution. Later on, the Vaidya\cite{Vaidya} metric corresponding to the exterior gravitational field of a stellar body describing an outgoing null fluid gave a tremendous impetus in this direction. Making use of the Vaidya\cite{Vaidya} solution, Santos\cite{Santos} proposed a procedure to obtain the description of the interior space-time of a spherically symmetric radially shrinking distribution of non-adiabatic fluid. Several stellar models (see for example, \cite{Bonnor,Oliveira1,Oliveira2,Oliveira3,Tikekar1,Mart,Maharaj,Herrera01,Herrera1,Herrera2,Herrera03,Herrera4,Herrera05,Herrera6,Chan1,Chan2,Chan3,Chan4,Pinheiro,Barreto,Ghezzi,Prisco1,Prisco2,Sharif1,Sharif2,Sharif3,Sharif4,Sharif5,Sharif6} and references therein) have been obtained and examined critically using Santos's approach. The procedure has been found to be useful to examine the impact of various factors such as shear, inhomogeneity, anisotropy, electromagnetic field and various dissipative processes on the evolution. In the absence of any established theory governing the nature of collapse, such investigations have been found to be very useful to understand the dynamic behaviour of a gravitationally collapsing system. 

The objective of the present work is to formulate a framework to study the nature of collapse of an inhomogeneously distributed anisotropic source. The dynamical behaviour of a gravitating system is expected to be influenced by density inhomogeneity and anisotropy. Amongst many other issues, these factors have been found to play crucial role in our theoretical understanding of the nature of singularity and its genericity (see \cite{Joshibook,Joshi01,Waugh,Joshi02,Singh,Goswami1,Chak} and references therein). Eardley and Smarr\cite{Eardley} have shown that inhomogeneous distributions may lead to a naked singularity in contrast to a homogeneous distribution where a black hole is more likely to be formed. For a spherically symmetric dust cloud, Mena {\em et al}\cite{Mena} have examined the role of inhomogeneity and anisotropy in Lamaitre-Tolman-Bondi collapse. Hererra {\em et al}\cite{Herrera03,Herrera05} have examined the role of various factors contributing to inhomogeneity of matter distribution, its evolution and development of anisotropy.  It has been observed that tidal forces tend to make a gravitating system more inhomogeneous which may lead to the formation of a naked singularity while for a homogeneously distributed body it is more likely to form a black hole. Similar observations may be found in \cite{Chan2} where it has been shown that if an initially static star undergoing collapse has a homogeneous distribution it ends up with the formation of a black hole. However, if an anisotropic star contains an inhomogeneous distribution of matter before the collapse sets in, then for a shearing model of collapse, the black hole is never formed as the condition for the formation of apparent horizon is never satisfied under such a condition\cite{Chan2}. 

Pressure anisotropy also plays a significant role in the construction of a dynamically evolving system. Elaborate discussions on the microscopic origin of anisotropic stresses in stellar bodies may be found in \cite{Herrera1},\cite{Bowers}. Amongst many other factors, electromagnetic field and shear have been found to contribute significantly to the generation of anisotropic stresses. The effect of charge on non-static models of gravitationally bound objects have been investigated by many authors\cite{Oliveira3,Tikekar1,Maharaj,Herrera1,Herrera03,Herrera4,Herrera05,Herrera6,Pinheiro,Barreto,Ghezzi,Prisco1,Prisco2,Sharif3,Sharif5,Sharif6,Chak,Beken,Rosales,Thiru} and references therein). Shear is also a generator of anisotropic stresses\cite{Prisco2}. Effects of shear and anisotropy in the dynamical behaviour of a self gravitating system have been analyzed by Govinder {\em et al}\cite{Govinder}. It has been observed that if a star has isotropic pressure before the collapse sets in, anisotropy may develop at a later stage due to the presence of shear\cite{Chan1,Chan3}.

In our work, we have developed a model describing a spherically symmetric inhomogeneous anisotropic fluid configuration radiating away its energy in the form of radial heat flux and shrinking in size as the collapse proceeds. In our construction, we have assumed that the collapsing star is charge neutral and the back ground space-time is shear-free. However, the energy-momentum tensor corresponding to the fluid distribution filling the interior of the collapsing star has been assumed to be anisotropic, in general. The model developed here is a generalization of an earlier model presented by Banerjee {\em et al}\cite{Baner} describing the collapse of a homogeneously distributed isotropic fluid. Sch$\ddot{a}$fer and Goenner\cite{Dirk} used the Banerjee {\em et al}\cite{Baner} model to investigate the collapse of a homogeneously distributed isotropic fluid configuration and also critically analyzed various features of the model parameters involving the background space-time. In the Sch$\ddot{a}$fer and Goenner\cite{Dirk} model, the collapse begins at time $t = - \infty$ with both infinite mass and radius and contracts to a point at time $t = 0$ without forming an event horizon. The horizon is never formed in this set up because the rate of collapse is counter balanced by the rate at which energy is dissipated to the exterior space-time of the star. 

In our construction, we have traced the evolution of the collapse on the background of space-time obtained by introducing an inhomogeneous perturbation in the Robertson-Walker space-time. Inhomogeneity in the geometry of the background space-time may be interpreted in the following way. We have considered the $3$-sub-space of the $4$-D manifold as having a geometry of a $3$-spheroid rather than a $3$-sphere as in \cite{Baner}. We have examined the impacts of inhomogeneous nature of background space-time and anisotropic stresses on the collapse by comparing the behaviour of physical parameters in our set up to the behaviour of corresponding parameters in Banerjee {\em et al}\cite{Baner} model, admissible as a special class in our model. We have also explored the thermal behaviour at the interior of the collapsing star by considering two different space-time backgrounds. We have not incorporated bulk viscosity and electromagnetic field in our model as these factors are likely to be absorbed into the anisotropic stresses. We have also considered a shear-free model so as to keep the governing field equations reasonably simple and tractable. Moreover, since the Sch$\ddot{a}$fer and Goenner\cite{Dirk} model was assumed to be shear-free, this simplifying assumption helps us to examine the impact of inhomogeneity directly by comparing our results to its homogeneous counterpart.

The paper has been organized as follows: In Section $2$, we have presented an inhomogeneous generalization of the Banerjee {\em et al}\cite{Baner} model by introducing a perturbation in the background space-time for a spherically symmetric anisotropic fluid undergoing non-adiabatic radiative collapse. Stipulating the boundary conditions across the surface separating the stellar configuration from the  Vaidya\cite{Vaidya} space-time describing its exterior filled with outgoing radiation, we have solved the surface equation which governs the evolution of the collapse. In Section $3$, we have examined the implications of  inhomogeneity on the collapse by comparing the evolutions of the physical quantities to the homogeneous model discussed in \cite{Dirk}. Bounds on the model paramters based on physical requirements and evolution of temperature have been analyzed in this Section. Finally, some concluding remarks have been made in Section $4$.

\section{Interior space-time}
The conformally flat space-time metric formulated by Maiti\cite{Maiti}, representing a spherically symmetric shear-free and rotation-free fluid with heat flux as source, has the form 
\begin{equation}
ds_{-}^2 = -\left[1+\frac{a(t)}{1+k(t)\frac{r^2}{4}}\right]^2 dt^2 + \frac{R^2(t)}{\left(1+k(t)\frac{r^2}{4}\right)^2}\left[dr^2+r^2d\Omega^2\right].\label{eq1}
\end{equation}
The line element (\ref{eq1}) reduces to Robertson-Walker metric for $a(t) =\dot{k} =0$.  Making use of the metric (\ref{eq1}), Banerjee {\em et al}\cite{Baner} presented a simple model for a collapsing body with outgoing radiation by assuming $a(\neq0)$ and $k$ as remaining constants during collapse. The collapsing matter in this set up was assumed to be a homogeneously distributed isotropic fluid. Various aspects of this model have been extensively examined by Sch$\ddot{a}$fer and Goenner\cite{Dirk} putting constraints on model parameters complying with various physical plausibility requirements. Following Sch$\ddot{a}$fer and Goenner\cite{Dirk}, we express the space-time metric of Banerjee {\em et al}\cite{Baner} model in standard coordinates as
\begin{equation}
ds_{-}^2 = -(C-\sqrt{1-kr^2})^2 dt^2 + R^2(t)\left[\frac{dr^2}{1-kr^2}+r^2d\Omega^2\right],\label{eq2}
\end{equation}
where, $C$ and $k(\neq0)$ are constants. 

To generalize the Banerjee {\em et al}\cite{Baner} model, we assume that the collapsing configuration is comprised of an inhomogeneous distribution of anisotropic fluid with its background space-time having the form
\begin{equation}
ds_{-}^2 = -A_0^2(r) dt^2 + R^2(t)\left[\frac{1+\lambda k r^2}{1-k r^2} dr^2 + r^2d\Omega^2\right]. \label{eq3}
\end{equation}
Here, the metric potential $A_0(r)$ and the scale factor $R(t)$ are undetermined metric functions and  $\lambda$  is  a parameter measuring departure from homogeneous geometry. The $t= constant$ hyper surface of the space-time (\ref{eq3}) has a geometry of a $3$-spheroid\cite{Vaidya2,Tikekar2,Knutsen,Maharaj2,Mukherjee} representing a perturbation from that of a $3$-sphere. The energy-momentum tensor of the fluid with anisotropy in pressure filling the interior of the collapsing body is written explicitly in the form 
\begin{equation}
T_{\alpha\beta} = (\rho + p_t) u_\alpha u_\beta + p_t g_{\alpha\beta} + (p_r - p_t)\chi_\alpha \chi_\beta +
q_\alpha u_\beta + q_\beta u_\alpha. \label{eq4}
\end{equation}
Here $\rho$ represents the energy density, $p_r$ is the radial pressure, $p_t$ is the tangential pressure, $\chi^{\alpha}$ is a unit space like four vector along the radial direction, $ u^{\alpha}$ is the 4-velocity of the fluid and $q^\alpha= q\delta^\alpha _r$ is the heat flux vector which is orthogonal to the velocity vector so that $q^\alpha u_\alpha = 0$.
Einstein's field equations in view of (\ref{eq3}) and (\ref{eq4}) lead to the following system of four independent equations
\begin{eqnarray}
8\pi \rho &=& \frac{1}{R^2}\left[\frac{1}{r^2}-\frac{1}{r^2 B_0^2}+\frac{2B'_0}{r B_0^3}\right]  + \frac{3\dot{R}^2}{A_0^2 R^2}, \label{eq5} \\
8\pi p_r &=& \frac{1}{ R^2}\left[-\frac{1}{r^2} + \frac{1}{B_0^2 r^2}+\frac{2 A'_0}{rA_0 B_0^2}\right] - \frac{1}{A_0^2}\left(\frac{\dot{R}^2}{R^2}+2\frac{\ddot{R}}{R}\right),\label{eq6}  \\ 
8\pi p_t &=& \frac{1}{R^2}\left[\frac{A''_0}{A_0 B_0^2}+\frac{A'_0}{rA_0 B_0^2}-\frac{B'_0}{rB_0^3}-\frac{A'_0 B'_0}{A_0 B_0^3}\right]-\frac{1}{A_0^2}\left[\frac{2\ddot{R}}{R}+\frac{\dot{R}^2}{R^2}\right],\label{eq7}  \\
8\pi q &=& - \frac{2A'_0\dot{R}}{A_0^2 B_0^2 R^3}, \label{eq8}
\end{eqnarray}
where,
\begin{equation}
B_0 = \sqrt{\left(\frac{1+\lambda k r^2}{1-k r^2}\right)}.\label{eq9}\\
\end{equation}
We have used in above the system of units rendering $G = c = 1$. Combining Eqs.~(\ref{eq6})-(\ref{eq7}), a time-independent differential equation of the form
\begin{equation}
\frac{A''_0}{A_0 B_0^2}-\frac{A'_0}{rA_0 B_0^2}-\frac{B'_0}{rB_0^3}-\frac{A'_0 B'_0}{A_0 B_0^3} - \frac{1}{B_0^2 r^2} +\frac{1}{r^2} - \delta(r) = 0,\label{eq10}
\end{equation}
may be obtained if it is assumed that the anisotropy evolves as
\begin{equation}
8\pi (p_t -p_r) = \Delta(r,t) = \frac{\delta(r)}{R^2(t)}.\label{eq11}
\end{equation}
By making a transformation $x^2=1-k r^2$ and using the expression for $B_0(r)$, Eq.~(\ref{eq10}) may be rewritten as
\begin{equation}
(1+\lambda k-\lambda k x^2)\frac{d^2 A_0}{dx^2} + \lambda k x \frac{d A_0}{dx} +\left(\lambda k(\lambda k+1)-\frac{(1+\lambda k-\lambda  k x^2)^2\delta(r)}{k(1-x^2)}\right)A_0= 0. \label{eq12}
\end{equation}
In terms of a new dependent variable
\begin{equation}
\Psi(x) = A_0(x)[1+\lambda k-\lambda k x^2]^{-1/4},\label{eq13}
\end{equation}
Eq.~(\ref{eq12}) assumes the form
\begin{equation}
\frac{d^2\Psi}{d x^2}+\left[\frac{2\lambda k(\lambda k+1)(2\lambda k+1)-(4\lambda k+7)\lambda^2 k^2 x^2}{4(1+\lambda k-\lambda  k x^2)^2}-\frac{(1+\lambda k-\lambda k x^2)\delta(r)}{k(1-x^2)}\right]\Psi = 0.\label{eq14}
\end{equation}
$\Psi$ can be determined from Eq.~(\ref{eq14}) if the nature of anisotropic parameter $\delta(r)$ is known. We shall examine the effect of inhomogeneity on the evolution of the collapse using the simple tractable solution
\begin{equation}
\Psi = C + D x,\label{eq15}
\end{equation}
of Eq.~(\ref{eq14}), obtained by stipulating
\begin{equation}
\delta (r) =  \frac{k(1-x^2)[2\lambda k(\lambda k+1)(2\lambda k+1)-(4\lambda k+7)\lambda^2 k^2 x^2]}{4(1+\lambda k-\lambda k x^2)^3}.\label{eq16}
\end{equation}
Note that the anisotropy parameter $\delta(r)$ is regular at all interior points of the configuration. In Eq.~(\ref{eq15}), $C$ and $D$ are arbitrary constants of integration. Combining Eqs.~(\ref{eq13}) and (\ref{eq15}), the metric function $A_0(r)$ is obtained as
\begin{equation}
A_0(r) = (1+\lambda k^2 r^2)^{1/4} (C +D\sqrt{1-k r^2}).\label{eq17}
\end{equation}
Consequently the space-time of the collapsing shear-free stellar body with anisotropic stresses is described by metric
\begin{equation}
ds_{-}^2 = -(1+\lambda k^2 r^2)^{1/2}(C +D\sqrt{1-k r^2})^2 dt^2 + R^2(t)\left[\frac{1+\lambda k r^2}{1-k r^2} dr^2 + r^2 d\Omega^2\right]. \label{eq18}
\end{equation}
The Banerjee {\em et al}\cite{Baner} model is a sub-class of (\ref{eq18}) which follows on setting $\lambda = 0$ and $D = -1$\cite{Dirk}. We shall examine the impact of the inhomogeneous nature of the background geometry due to presence of $\lambda$ without any loss of generality by stipulating $D = -1$ .

\subsection{Determination of $R(t)$}
The evolution of the collapse in the stellar body with interior space-time metric (\ref{eq18}) is governed by the function $R(t)$ which has role of a scale factor in the process. It is determined by the boundary conditions across the boundary surface $\Sigma$ of the configuration as it shrinks in size. The space-time outside the collapsing body will be appropriately described by the Vaidya\cite{Vaidya} metric
\begin{equation}
ds_{+}^2 = -\left(1-\frac{2m(v)}{\bar{r}}\right)dv^2 - 2dvd\bar{r} + \bar{r}^2d \Omega^2.\label{eq19}
\end{equation}
Here $v$ denotes retarded time and $m(v)$ represents the total mass of the collapsing star which is expected to be a function of the retarded time $v$. Following the method presented by Santos\cite{Santos},  we write the matching conditions linking smoothly the interior and the exterior space-times across the boundary $3$-space-time of the evolving system as
\begin{eqnarray}
m(v) &=& \frac{(r R(t))_{\Sigma}}{2}\left[1 -\frac{1}{B_0^2}+\left(\frac{r\dot{R}}{A_0}\right)^2\right]_{\Sigma},\label{eq20}\\
(p_r)_{\Sigma} &=& (q R(t)B_0)_{\Sigma}.\label{eq21}
\end{eqnarray}
From (\ref{eq20}) it follows that the mass of matter enclosed within the boundary surface $r < r_\Sigma$ will be
\begin{equation}
m(r,t) = \frac{rR(t)}{2}\left[1 -\frac{1}{B_0^2}+\left(\frac{r\dot{R}}{A_0}\right)^2\right].\label{eq22}
\end{equation}

Eqs.~(\ref{eq6}), (\ref{eq8}) and the boundary condition (\ref{eq21}) at $r=r_{\Sigma}$ lead to
\begin{equation}
\ddot{R}R + \frac{1}{2}\dot{R}^2 -\alpha\dot{R} +\beta = 0, \label{eq23}
\end{equation}
where, $\alpha$ and $\beta$ are constants given by
\begin{eqnarray}
\alpha &=& \frac{C\lambda k^2r_{\Sigma}\sqrt{1-k r_{\Sigma}^2} + kr_{\Sigma}(2-\lambda k + 3\lambda k ^2 r_{\Sigma}^2)}{2\sqrt{1+\lambda k r_{\Sigma}^2}(1+\lambda k^2 r_{\Sigma}^2)^{3/4}},\label{eq24}\\
\beta &=& \frac{k(C-\sqrt{1-k r_{\Sigma}^2})}{2(1+\lambda k r_{\Sigma}^2)\sqrt{1+\lambda k^2 r_{\Sigma}^2}}\left[C(1+\lambda -\lambda k +2 \lambda k^2 r_{\Sigma}^2+\lambda^2 k^2 r_{\Sigma}^2)\right.\nonumber\\
&&\left.-\frac{(3+\lambda-\lambda k -\lambda k r_{\Sigma}^2+5\lambda k^2 r_{\Sigma}^2+\lambda^2 k^2 r_{\Sigma}^2-3k r_{\Sigma}^2-\lambda^2 k^3 r_{\Sigma}^4-4\lambda k^3 r_{\Sigma}^4)}{\sqrt{1-k r_{\Sigma}^2}}\right].\label{eq25}
\end{eqnarray} 
A simple solution of Eq.~(\ref{eq23}) will then be\cite{Dirk}
\begin{equation}
R(t) = n t,\label{eq26}
\end{equation}
where,
\begin{equation}
n= \alpha \pm \sqrt{\alpha^2 -2\beta}.\label{eq27}
\end{equation}
For collapsing configurations we must have $\dot{R} < 0$. Hence $n$ should be negative implying
\begin{equation}
n= \alpha - \sqrt{\alpha^2 -2\beta}.\label{eq28}
\end{equation}
The constants $C$, $k$ and $\lambda$ should be chosen to ensure $\alpha ^2 > 2\beta$ and $n<0$.  

\section{Physical analysis}
Einstein's field equations lead to the following explicit expression for the dynamical variables of matter density, the radial and transverse pressures and the heat flux parameter associated with the collapsing stellar structure
\begin{eqnarray}
8\pi \rho &=& \frac{k(\lambda +1)(3+\lambda k r^2)}{n^2 t^2 (1+\lambda k r^2)^2}+\frac{3}{t^2(1+\lambda k^2 r^2)^{1/2} (C -\sqrt{1-k r^2})^2},\label{eq29}\\
8\pi p_r &=& 8\pi \rho - \frac{4}{t^2(1+\lambda k^2 r^2)^{1/2} (C -\sqrt{1-k r^2})^2},\label{eq30}\\
8\pi p_t &=& 8\pi p_r +\frac{k^2r^2\left[2\lambda k(1+\lambda k)(1+2\lambda k)-\lambda^2 k^2(1-k r^2)(4\lambda k +7)\right]}{4n^2 t^2(1+\lambda k^2 r^2)^3},\label{eq31}\\
8\pi q &=& -\frac{ r\sqrt{1-k r^2}\left[C \lambda k \sqrt{1-k r^2} +(2-\lambda k +3\lambda k^2 r^2)\right]}{n^2 t^3 (1+\lambda k r^2)(C-\sqrt{1-k r^2})^2(1+\lambda k^2 r^2)^{5/4}}.\label{eq32}
\end{eqnarray}
These physical parameters have zero values at $t\rightarrow -\infty$; the density and the two pressures evolve with time as $\sim 1/t^2$ while the heat flux evolves as $\sim 1/t^3$. The proper radius $r_p = [R(t)r]_{\Sigma} = ntr_{\Sigma}$ is infinite when collapse begins, positive at any later instant $t$ and shrinks to zero at $t=0$. The mass at any instant $t$ within the boundary radius $r_{\Sigma}$ has expression
\begin{equation}
m(v) \stackrel{\Sigma}{=} m(r_{\Sigma},t) = \frac{nr_{\Sigma}t}{2}\left[\frac{(1+\lambda)k r_{\Sigma}^2}{(1+\lambda k r_{\Sigma}^2)}+\frac{n^2 r_{\Sigma}^2}{(C-\sqrt{1-k r_{\Sigma}^2})^2(1+\lambda k^2 r_{\Sigma}^2)^{1/2}}\right],\label{eq33}
\end{equation}
which shows that the collapse begins with an infinite mass and size of the configuration at $t\rightarrow -\infty$ and it evaporates completely as the epoch $t=0$ approaches. As in \cite{Banerjee2}, the collapsing matter radiates all the mass energy when the singularity is reached as proper boundary radius of the configuration  $r_p $ =  $[rR(t)]_{\Sigma} = 0$ at  $t=0$. 

\subsection{Nature of singularity}
To understand the nature of singularity, in general, it is necessary to verify the existence of radial null geodesics emerging from the singularity. The existence of a naked singularity is characterized by the absence of apparent horizon as the collapse approaches the central singularity $r=0$. However, such a test is not required in our model since the ratio of $(2m(r,t)/{rR})_{\Sigma}$ is independent of time. One can always suitably choose the model parameters so that the ratio $(2m(r,t)/{rR})_{\Sigma} < 1$. The function $1-[2m(r,t)/(rR)]_{\Sigma}$ will then remain positive throughout the collapse process which indicates that the collapse would terminate into a naked singularity. The Ricci curvature, in our model, turns out to be
\begin{equation}
\Re = \frac{6r^2n^2 B_0^2 -2A_0^2(B_0-2rB_0')+2rA_0[A_0'B_0'-B_0(2A_0'+rA_0'')]}{r^2A_0^2B_0^2n^2t^2},\label{cureq1}
\end{equation}
which show that the curvature diverges as $1/t^2$. This suggests that it is a weak curvature singularity.

\subsection{Energy conditions}
A physically reasonable solution should satisfy certain energy conditions, namely, (1) the null energy condition ($\rho \geq 0$); (2) the weak energy condition ($\rho - p_r \geq 0$, $\rho - p_t \geq 0$); and (3) the dominant energy condition ($\rho - p_r - 2p_t \geq 0$).

From Eq.~(\ref{eq29})-(\ref{eq31}), it is easy to show that $\rho,~p_r,~p_t > 0$ and $\rho',~p_r',~p_t' < 0$, if the conditions $\lambda > 0$ and $ k > 0$ are satisfied simultaneously. Eq.~(\ref{eq30}) shows that $\rho > p_r$. Note that at $r=0$, $p_r=p_t$ and both $p_r$ and $p_t$ decrease radially outward. We, therefore, may conclude that the null and weak energy conditions are satisfied in our model. It is difficult to verify the dominant energy condition unless numerical techniques are invoked. However, we note that by setting $R(t) = 1$ in our model, it is possible obtain an initial static configuration. The $t= constant$ hypersurface of the initial static configuration describes a space-time geometry which is spheroidal in nature. The corresponding static configurations have been found to comply with the energy conditions mentioned above (see \cite{Shib} and references therein). Since $\rho$, $p_r$ and $p_t$ evolve as $1/t^2$, it is expected that the fulfillment of the energy conditions will be sustained throughout the collapse.

\subsection{Bounds on model parameters}
\begin{figure}
\includegraphics[width=0.6\textwidth]{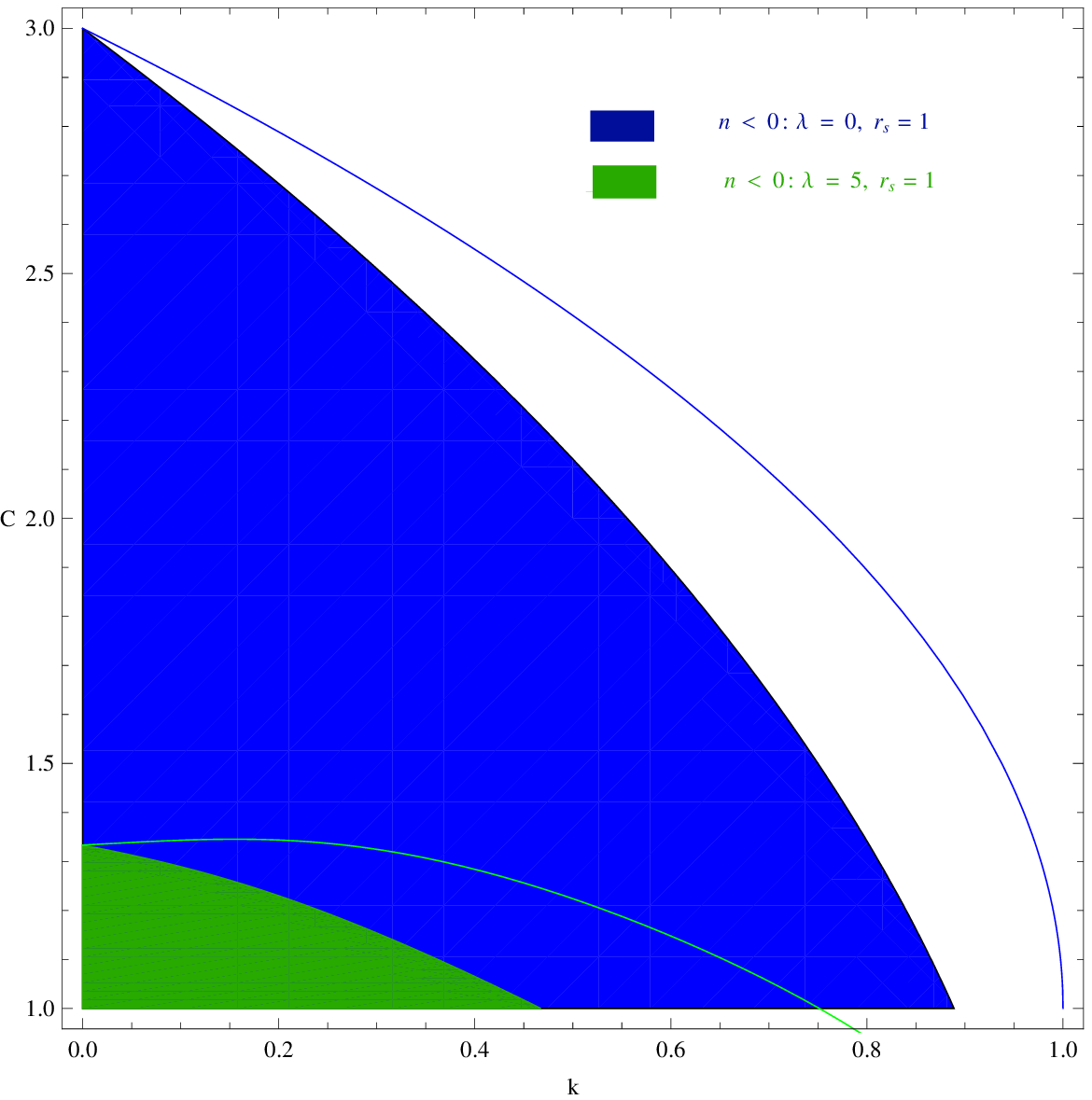}
\caption{Range for the parameters $k$ and $C$ for $\lambda=0$ (blue shaded region which includes the green shaded region) and $\lambda =5$ (green shaded region) obtained by imposing the condition $n < 0$. For $\lambda =0$ and $5$, the blue and green curves respectively give the bounds on the model parameters obtained from the condition (c).}
\label{fig:1}
\end{figure}

The rate of expansion, in our model, has the explicit expression
\begin{equation}
\Theta = u^{\alpha}_{;\alpha}= \frac{3\dot{R}}{A_0 R} = \frac{3}{t(1+\lambda k^2 r^2)^{1/4} (C - \sqrt{1-k r^2})}, \label{eq34}
\end{equation}
and should be negative since our model describes a contracting body. Therefore, during the whole collapse process from time $t = - \infty$ to $t = 0$, we must have
\begin{equation}
(1+\lambda k^2 r^2)^{1/4} (C - \sqrt{1-k r^2}) > 0. \label{eq35}
\end{equation}
Eq.~(\ref{eq35}) implies a lower bound on $C$ such that $C > 1$. Eq.~(\ref{eq35}) also follows from the requirement  $A_0(r) > 0$.  The constraints on the model parameters obtained in view of physical viability are specified below:
\begin{description}
\item[(a)] From Eq.~(\ref{eq29}), the central density is obtained as
\begin{equation}
\rho_c = \frac{3k(\lambda+1)}{n^2t^2} +\frac{3}{t^2(C-1)^2}.\nonumber
\end{equation}
Therefore, for a positive energy density, we must have $k > 0$. Moreover, the requirements $\rho,~p_r,~p_t > 0$ and $\rho',~p_r',~p_t' < 0$, demand that $\lambda > 0$. 

\item[(b)] For $ 0 \leq r \leq r_{\Sigma}$, the factor $\sqrt{1-kr^2}$ appearing in the metric function $A_0(r)$ will be real if $ 0 < k \leq 1/r_{\Sigma}^2$.

\item[(c)] As discussed earlier, the lower bound on $C$ is given by $C > 1$. The upper bound on $C$ may be obtained by noting that $\alpha^2 > 2\beta$ which ensures that $n$ in Eq.~(\ref{eq28}) remains real. Thus, bounds on $C$ may be expressed as
\begin{equation}
 1< C < \frac{8\sqrt{1-kr_{\Sigma}^2}+4\sqrt{1+\lambda k r_{\Sigma}^2}(1+\lambda k^2 r_{\Sigma}^2)^2+a\lambda\sqrt{1-k r_{\Sigma}^2}}{4+b\lambda},\nonumber
\end{equation}
where,
\begin{eqnarray}
a &=& 4-4k+22k^2 r_{\Sigma}^2+8\lambda k^2 r_{\Sigma}^2-5\lambda k^3 r_{\Sigma}^2+15\lambda k^4 r_{\Sigma}^2+4\lambda^2 k^4r_{\Sigma}^4, \nonumber \\
b &=& 4-4k+12k^2r_{\Sigma}^2+8\lambda k^2r_{\Sigma}^2-5\lambda k^3r_{\Sigma}^2+9\lambda k^4r_{\Sigma}^4+4\lambda^2 k^4r_{\Sigma}^4. \nonumber
\end{eqnarray}

\item[(d)] For collapsing configurations we must have $\dot{R} < 0$ which implies that $n < 0$.
\end{description}
Though, more stringent constraints on the model parameters can not be ruled out, the constraint equations obtained previously by Sch$\ddot{a}$fer and Goenner\cite{Dirk} may be regained by stipulating $\lambda =0$ in this model. The parameter $\lambda$ will take on value such that all the above mentioned conditions are satisfied simultaneously. In Fig.~(\ref{fig:1}), we have shown the range of values for $k$ and $C$ consistent with the constraints which can ensure regular behaviour of the physical parameters. Note that the condition (d) yields a more stringent bound on the parameters as compared to the condition (c).  

\begin{figure}
\includegraphics[width=0.6\textwidth]{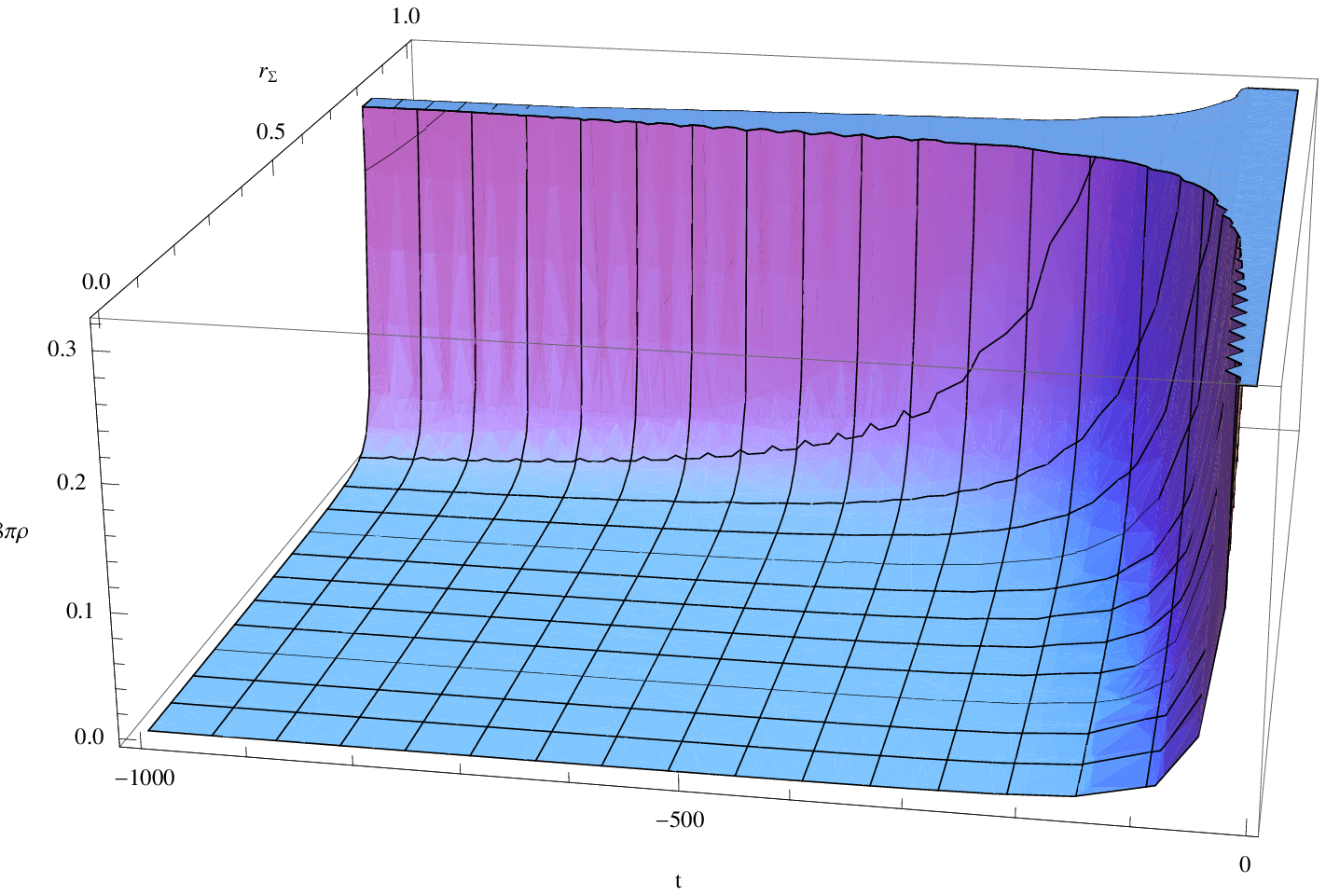}
\caption{Evolution of the energy density ($8\pi \rho$) at the centre ($r=0$) for an inhomogeneous and  anisotropic  distribution ($\lambda=5$, $k=0.2$, $C=1.2$).}
\label{fig:2}
\end{figure}

\begin{figure}
\includegraphics[width=0.6\textwidth]{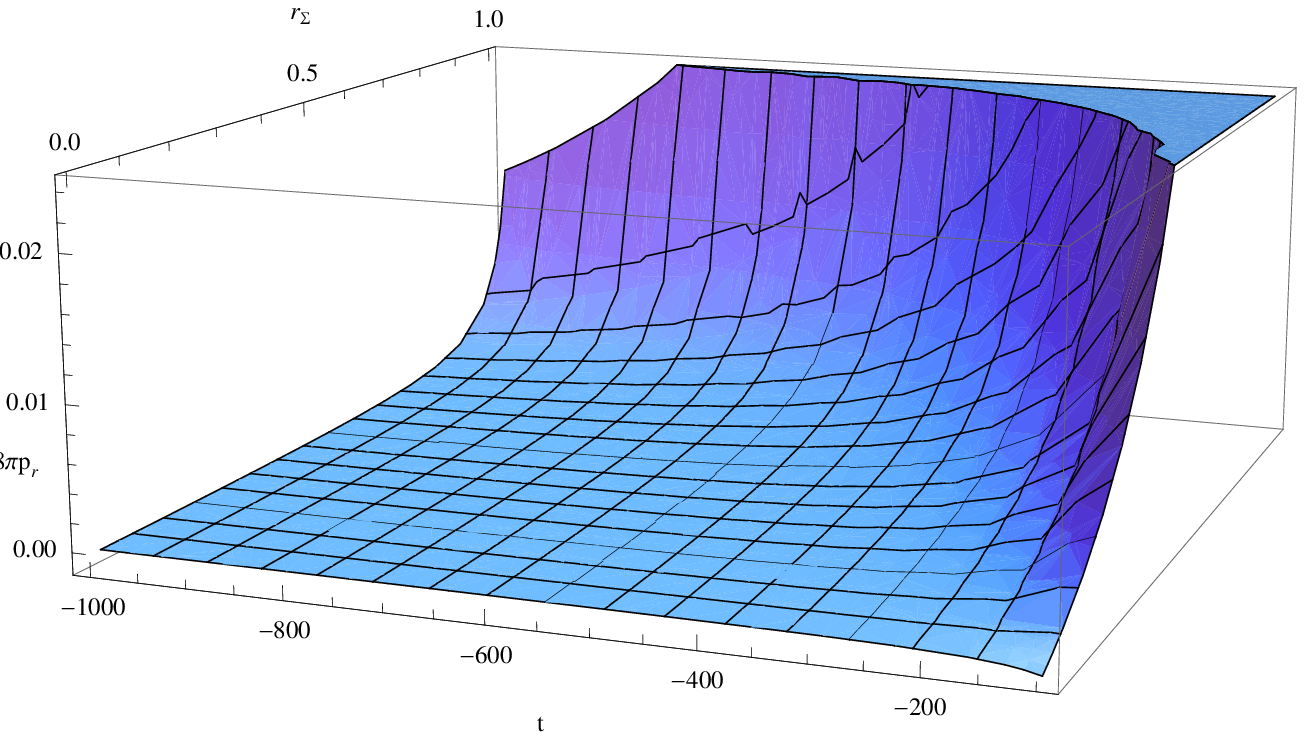}
\caption{Evolution of the radial pressure ($8\pi p_r$) at the centre ($r=0$) for an inhomogeneous  and  anisotropic  distribution ($\lambda=5$, $k=0.2$, $C=1.2$).}
\label{fig:3}
\end{figure}

\begin{figure}
\includegraphics[width=0.6\textwidth]{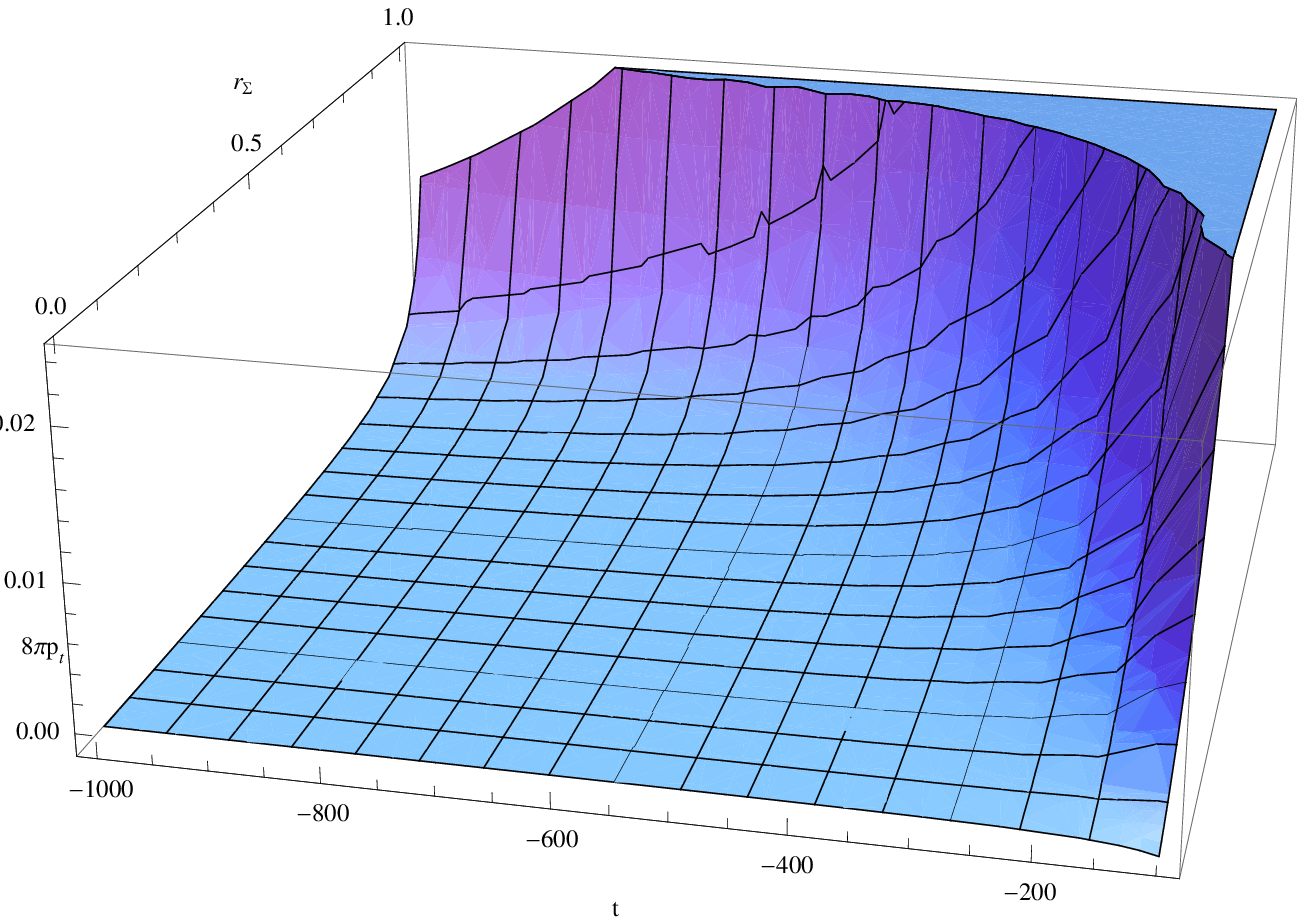}
\caption{Evolution of the transverse pressure ($8\pi p_t$) at the centre ($r=0$) for an inhomogeneous and  anisotropic  distribution ($\lambda=5$, $k=0.2$, $C=1.2$).}
\label{fig:4}
\end{figure}

\begin{figure}
\includegraphics[width=0.6\textwidth]{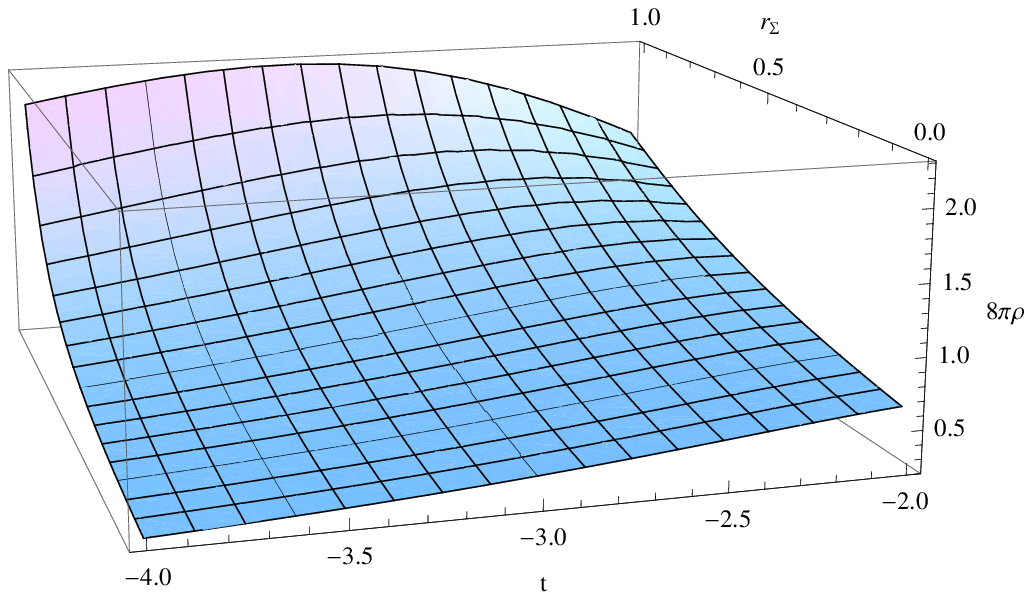}
\caption{Evolution of the energy density ($8\pi \rho$) at the centre ($r=0$) for a homogeneous and isotropic  distribution ($\lambda=0$, $k=0.2$, $C=2.5$).}
\label{fig:5}
\end{figure}

\begin{figure}
\includegraphics[width=0.6\textwidth]{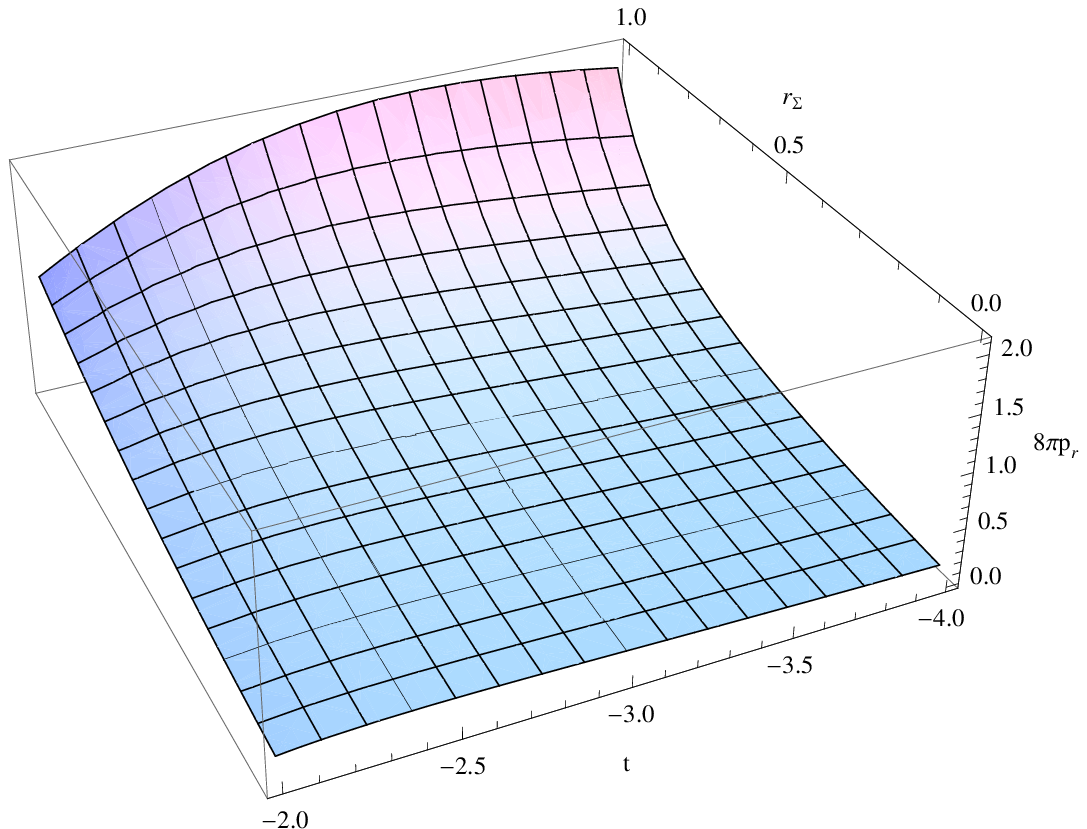}
\caption{Evolution of the radial pressure ($8\pi p_r$) at the centre ($r=0$) for a homogeneous and isotropic distribution ($\lambda=0$, $k=0.2$, $C=2.5$).}
\label{fig:6}
\end{figure}

\begin{figure}
\includegraphics[width=0.6\textwidth]{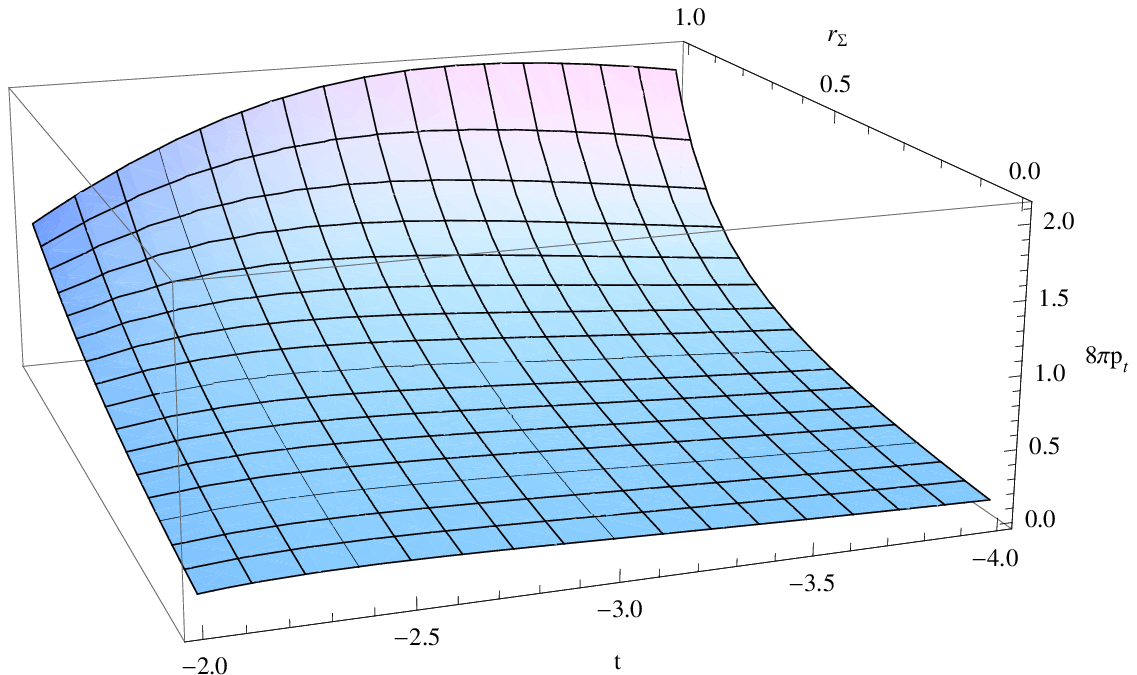}
\caption{Evolution of the transverse pressure ($8\pi p_t$) at the centre ($r=0$) for a homogeneous and isotropic  distribution ($\lambda=0$, $k=0.2$, $C=2.5$).}
\label{fig:7}
\end{figure}

We have checked the behaviour of the physical parameters like energy density, radial and tangential pressures by assuming a set of values of $\lambda$, $k$ and $C$ consistent with the above constraints. Fig.~(\ref{fig:2})-(\ref{fig:7}) describe the behaviour of various physical parameters and indicate that they all comply with the requirement of the regularity.  In the figures, we have assumed that the collapsing object has an initial radius $r_{\Sigma}(t\rightarrow -\infty) = r_s = 1$ as in \cite{Dirk}. The proper radius  $r_p = R(t)r_{\Sigma} = ntr_{\Sigma}$, shrinks to zero at $t = 0$ as $R(t) = 0$ at $t = 0$. Therefore, a choice of $r_s$ is possible for convenience and will not have any impact on the nature of collapse. The choice $r_s =1$ has the effect of changing the length measuring scale only. The bounds on the model parameters are valid for all $0 \leq r_{\Sigma} \leq \infty$ including  $r_{\Sigma}(t\rightarrow -\infty) = r_s = 1$.

\begin{figure}
\includegraphics[width=0.6\textwidth]{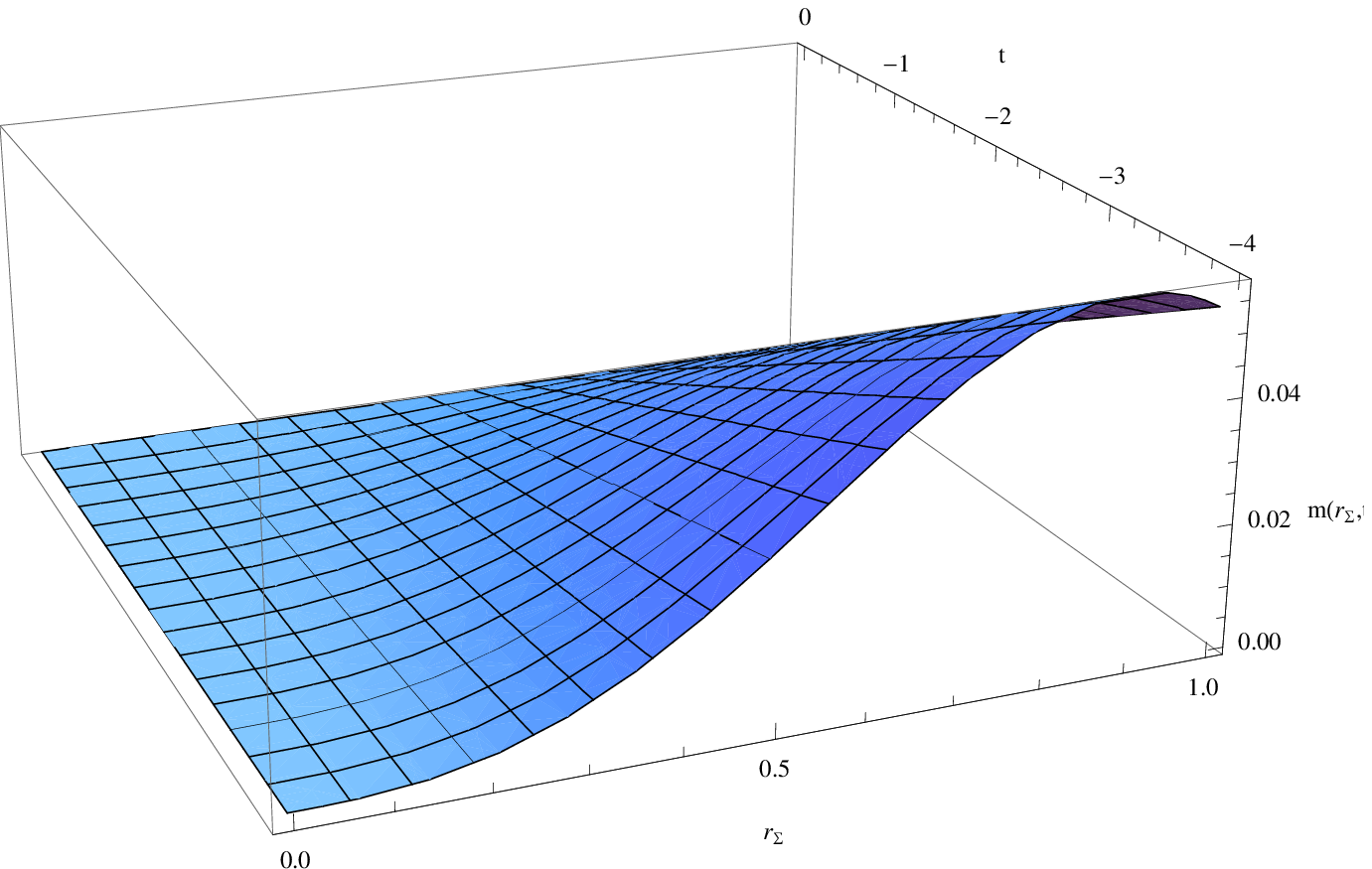}
\caption{Evolution of the mass function $m(r_{\Sigma},t)$ for a homogeneous and  isotropic distribution ($\lambda=0$, $k=0.2$, $C=2.5$).}
\label{fig:8}
\end{figure}

\begin{figure}
\includegraphics[width=0.8\textwidth]{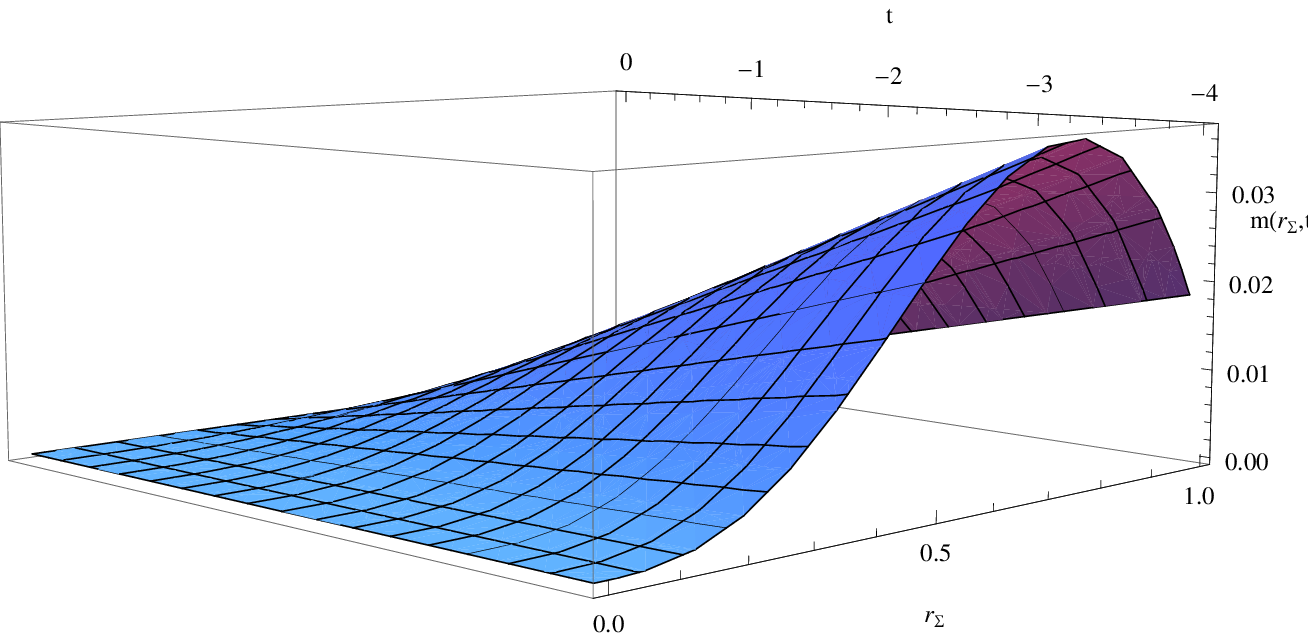}
\caption{Evolution of the mass function $m(r_{\Sigma},t)$ for an inhomogeneous and anisotropic  distribution ($\lambda=5$, $k=0.2$, $C=1.2$).}
\label{fig:9}

\end{figure}

\begin{figure}
\includegraphics[width=0.6\textwidth]{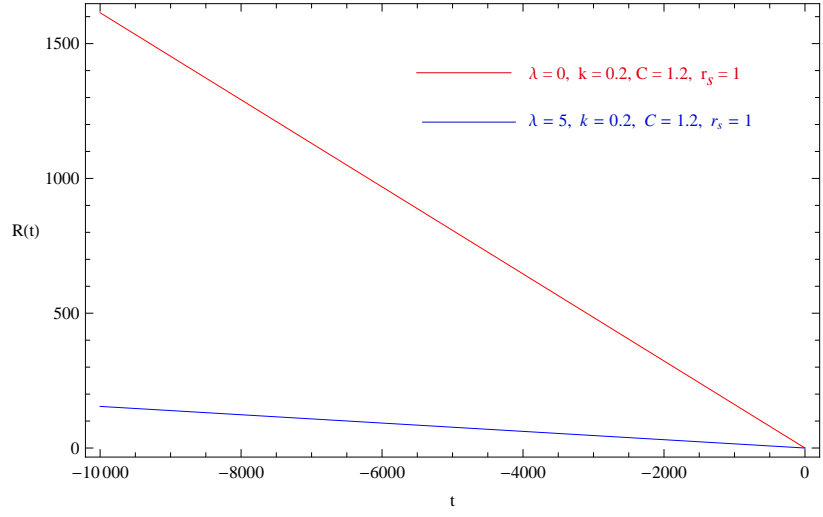}
\caption{Evolution of Scale factor $R(t)$ for homogeneous ($\lambda=0$) and inhomogeneous ($\lambda \neq 0$) distributions.}
\label{fig:10}
\end{figure}

In Fig.~(\ref{fig:2})-(\ref{fig:4}), we have shown the evolution of the central values of the energy density, radial and tangential pressures, respectively, by assuming a set of values consistent with the constraints (we have assumed $\lambda=5$, $k=0.2$, $C=1.2$) for an inhomogeneous anisotropic distribution.  Setting $\lambda=0$, $k=0.2$, $C=2.5$, behaviour of the corresponding parameters for a homogeneous isotropic distribution have been shown in Fig.~(\ref{fig:5})-(\ref{fig:7}). The plots clearly indicate that the physical parameters start with zero value at $t \rightarrow -\infty$ and increase till the singularity is reached at $t=0$ and $r_{\Sigma} =0$. Since anisotropy vanishes at the centre, evolution of the radial and tangential pressures show identical behaviour though explicit expressions for $p_r$ and $p_t$ in Eqs.~(\ref{eq30})-(\ref{eq31}) clearly indicate their inequality for $r\neq0$. If $\lambda=0$, $p_r =p_t$ at all interior points of the star which implies that pressure isotropy condition can be regained by setting $\lambda =0$.  In Fig.~(\ref{fig:8}) -(\ref{fig:9}), we have shown the variations of the total mass $m(r_{\Sigma},t)$ for $\lambda=0$ and $\lambda=5$, respectively. Evolution of the scale factor $R(t)$ has been shown in Fig.~(\ref{fig:10}) for two different values of $\lambda$($= 0,~5$). The $\lambda=0$ case corresponds to the Sch$\ddot{a}$fer and Goenner\cite{Dirk} model describing the collapse of a homogeneous fluid model with isotropic fluid pressure.

\subsection{Thermal behaviour}
\begin{figure}
\includegraphics[width=0.6\textwidth]{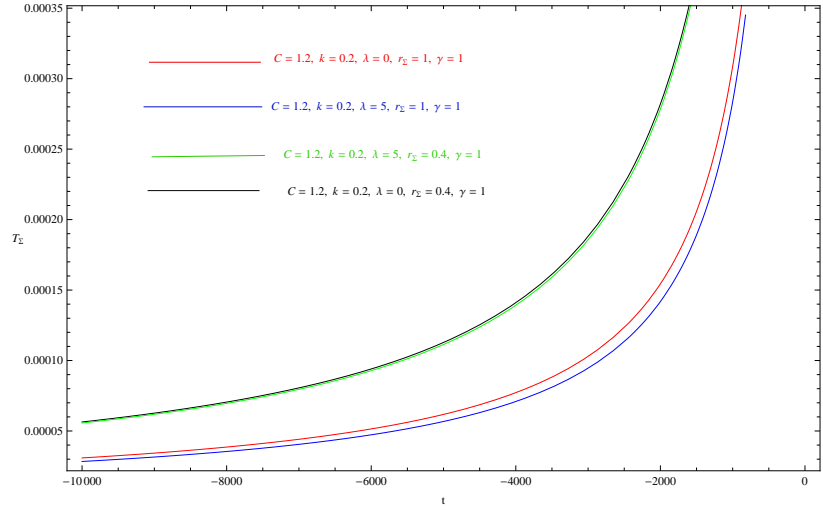}
\caption{Time evolution of the surface temperature $T_{\Sigma}$ for homogeneous ($\lambda = 0$) and inhomogeneous ($\lambda = 5$) distributions.}
\label{fig:11}
\end{figure}

Let us now investigate the evolution of temperature of the collapsing star. In view of extended irreversible thermodynamics, the relativistic Maxwell-Cattaneo relation for temperature governing the heat transport within the collapsing matter in the truncated Israel-Stewart theory\cite{Israel,Maartens,Martinez,Sarwe} has the form
\begin{equation}
\tau(g^{\alpha\beta}+u^{\alpha}u^{\beta})u^{\delta}q_{\beta;\delta} + q^{\alpha} = -\kappa(g^{\alpha\beta}+u^{\alpha}u^{\beta})[T_{,\beta}+T\dot{u_{\beta}}],\label{eqt1}
\end{equation}
where $\kappa (\geq 0)$ is the thermal conductivity and $\tau (\geq 0)$ is the relaxation time. In view of the line element (\ref{eq18}), Eq.~(\ref{eqt1}) reduces to
\begin{equation}
\tau\frac{d}{dt}(qRB_0) + qRA_0B_0 = -\kappa \frac{1}{RB_0}\frac{d}{dr}(A_0T)\label{eqt2}
\end{equation}
The relativistic Fourier heat transport equation may be obtained by setting $\tau = 0$ in (\ref{eqt2}). To get a simple estimate of the temperature evolution, we set $\tau=0$ in our calculations. For $\tau=0$, combining Eqs.~(\ref{eq8}) and (\ref{eqt2}), we obtain
\begin{equation}
8\pi\kappa(A_0T)' = \frac{2A_0'\dot{R}}{A_0R},\label{eqt3}
\end{equation}
Following an earlier treatment\cite{Sarwe}, we choose the thermal conductivity parameter in the form $\kappa = \gamma T^{\omega}$, where $\gamma$ and $\omega$ are constants. Eq.~(\ref{eqt3}), then,  yields
\begin{equation}
8\pi\gamma(A_0T)' = \frac{2A_0'}{A_0}\frac{T^{-\omega}}{t},\label{eqt4}
\end{equation}
where we have used Eq.~(\ref{eq26}). Integrating the above equation, we get
\begin{equation}
8\pi\gamma A_0T^{1+\omega} = \frac{2 Ln [A_0]}{t}.\label{eqt5}
\end{equation}
The surface temperature at any instant for the special case $(\omega = 0)$ in our model thus turns out to be
\begin{equation}
T(r_{\Sigma},t) = \frac{Ln[(1+\lambda k^2 r_{\Sigma}^2)^{1/4} (C -D\sqrt{1-k r_{\Sigma}^2})]}{4\pi\gamma t(1+\lambda k^2 r_{\Sigma}^2)^{1/4} (C -D\sqrt{1-k r_{\Sigma}^2})}.\label{eqt6}
\end{equation}
Obviously, the surface temperature has zero value at $t \rightarrow -\infty$ and it evolves as $1/t$. Time evolution of the surface temperature for inhomogeneous ($\lambda = 5$) and homogeneous ($\lambda = 0$) distributions has been shown in Fig.~(\ref{fig:11}). An interesting feature of our model is that the surface temperature for the inhomogeneous, anisotropic distribution appears to be less than that of its homogeneous, isotropic counterpart. However, evolution of the temperature shows identical behaviour in both the cases.

\section{Discussion}
We have generalized the Banerjee {\em et al}\cite{Baner} model and examined the impacts of inhomogeneous nature of background space-time and anisotropic stresses on the collapse by comparing the behaviour of physical parameters in our set up to the behaviour of corresponding parameters of a collapsing homogeneous, isotropic stellar configuration of Sch$\ddot{a}$fer and Goenner\cite{Dirk} model, admissible as a particular class. We have considered two cases setting the parameter $\lambda = 0,~5,$ for studying the evolution of the collapse using numerical procedures. The parameter $\lambda$ is a geometrical parameter which is a measure of inhomogeneity. We have derived new bounds on the model parameters and the bounds on the model parameters found in \cite{Dirk} may be regained simply by setting $\lambda=0$. 

Though, in general, the physical parameters evolve differently for two different back ground space-times, an inhomogeneous perturbation of the background space-time of the collapsing stellar configuration seems to have very little impact on the gross features of the evolving system. Apparently departure from homogeneity does not have any drastic impact on the evolution of collapse except for a change in the collapse rate by a factor $(1+\lambda k^2 r^2)^{1/4}$. The possibilities of an in-depth analysis in different set up giving conflicting results cannot be ruled out.

\begin{acknowledgements}
We would like to thank the anonymous referees for useful suggestions. RS gratefully acknowledges support from IUCAA, Pune, India, under Visiting Research Associateship Programme. RT would like to thank the UGC of India for Award of Emeritus Fellowship and IUCAA, Pune, for hospitality and facilities for work.
\end{acknowledgements}


\end{document}